\RequirePackage{fix-cm}
\documentclass[twocolumn,epjc3]{svjour3}  
\smartqed  
\RequirePackage{color,graphicx,url}
\usepackage{amsfonts,amsmath,amssymb,bm,xspace}
\usepackage{lipsum}
\usepackage[T1]{fontenc} 
\usepackage{epsfig}  
\usepackage{graphicx}  
\usepackage{amssymb}
\usepackage{amsmath}
\usepackage{pifont}
\usepackage[english]{babel}
\usepackage{color}  
\usepackage{multirow}
\definecolor{darkgreen}{rgb}{0.2,0.5, 0.2}
\usepackage{lipsum}
\usepackage{mathtools}
\usepackage{cuted}
\usepackage{dcolumn,booktabs}
\usepackage{ulem}
\newcolumntype{d}[1]{D{.}{.}{#1}}
\usepackage{cite} 
\usepackage{soul}
\usepackage{xcolor}
\newcommand{\sat}{\mathrm{sat}}

\def\ga{\,\,\raise0.14em\hbox{$>$}\kern-0.76em\lower0.28em\hbox  
{$\sim$}\,\,}  
\def\la{\,\,\raise0.14em\hbox{$<$}\kern-0.76em\lower0.28em\hbox  
{$\sim$}\,\,}  
\journalname{Eur. Phys. J. A}
\begin{document}

\title{Skyrme-Hartree-Fock-Bogoliubov mass models on a 3D mesh: IV. Improved description of the isospin dependence of pairing}


\author{Guilherme Grams\thanksref{e1,addr1,addr2}
        \and
        Nikolai N.~Shchechilin\thanksref{e2,addr1,addr2} 
        \and
        Adrian Sanchez-Fernandez\thanksref{addr1,addr2} 
        \and
        Wouter Ryssens\thanksref{addr1,addr2} 
        \and
        Nicolas Chamel\thanksref{addr1,addr2} 
        \and
        Stephane Goriely\thanksref{addr1,addr2} 
}
\thankstext{e1}{e-mail: guilherme.grams@ulb.be}
\thankstext{e2}{e-mail: nikolai.shchechilin@ulb.be}

\institute{Institut d’Astronomie et d’Astrophysique, Université Libre de Bruxelles, Brussels, Belgium \label{addr1} \and Brussels Laboratory of the Universe - BLU-ULB \label{addr2}
}

\date{Received: date / Accepted: date}
\maketitle

\begin{abstract}

Providing reliable data on the properties of atomic nuclei and infinite nuclear matter to astrophysical applications remains extremely challenging, especially when treating both properties coherently within the same framework. Methods based on energy density functionals (EDFs) enable manageable calculations of nuclear structure throughout the entire nuclear chart and of the properties of infinite nuclear matter across a wide range of densities and asymmetries. 
To address these challenges, we present BSkG4, the latest Brussels-Skyrme-on-a-Grid model. It is based on an EDF of the extended Skyrme type with terms that are both momentum and density-dependent, and refines the treatment of $^1S_0$ nucleon pairing gaps in asymmetric nuclear matter
as inspired  
by more advanced many-body calculations. 
The newest model maintains the accuracy of earlier BSkGs for known atomic masses, radii and fission barriers with rms deviations of 0.633 MeV w.r.t. 2457 atomic masses,  0.0246 fm w.r.t. 810 charge radii, and 0.36 MeV w.r.t 45 primary fission barriers of actinides. It also improves some specific pairing-related properties, such as the $^1S_0$ pairing gaps in asymmetric nuclear matter, neutron separation energies, $Q_\beta$ values, and moments of inertia of finite nuclei. 
This improvement is particularly relevant for describing the $r$-process nucleosynthesis as well as various astrophysical phenomena related to the rotational evolution of neutron stars, their oscillations, and their cooling.
\end{abstract}
%

\keywords{energy density functional \and pairing \and neutron star \and superfluidity}
%

\section{Introduction}
\label{intro}

Nuclear physics predictions are essential to many astrophysics applications, 
including simulations of the rapid neutron capture process (or $r$-process) and the modeling of neutron stars (NS). 
However, accurately describing the properties of the thousands of atomic nuclei 
involved in the $r$-process or of nuclear matter remains 
an arduous task, further complicated by the wide range of densities and neutron-proton 
asymmetries found in astrophysical environments~\cite{Arnould20,Oertel17,Burgio18}.
The challenge lies in developing a model that can simultaneously describe 
the equation of state (EoS) of dense matter in NS~\cite{Oertel17} and  
nuclear structure properties (such as binding energies, nuclear level densities, fission probabilities, and 
photon or beta strength functions) that are crucial for modeling reactions and decays~\cite{Arnould20}.


Methods based on energy density functionals (EDFs) provide an effective tool for supplying 
nuclear data to astrophysics, offering a quantum description of nuclear systems in terms of 
neutrons and protons~\cite{Bender03}. 
By connecting nucleonic densities to an effective nucleon-nucleon interaction, EDFs allow for 
tractable calculations of nuclear structure across the entire nuclear chart and of infinite 
nuclear matter (INM) properties for a wide range of densities. This makes EDFs a powerful 
framework for advancing global nuclear models relevant to astrophysics.
Within this framework, we have been developing the Brussels-Skyrme-on-a-Grid (BSkG)~\cite{Scamps21,Ryssens22,Grams23} 
series of global nuclear structure models.
The BSkG family is based on EDF of extended Skyrme type with both momentum- and density-dependent $t_4$ and 
$t_5$ terms, and uses the Hartree-Fock-Bogoliubov (HFB) 
approximation to address the many-body problem. These models aim to provide a comprehensive and as 
microscopic as possible description of nuclear structure and INM, with a specific focus on astrophysical 
applications. The BSkG series exploits the powerful concept of symmetry breaking, allowing for complex 
nuclear shapes, such as triaxial and octupole deformations at ground-state~\cite{Grams23} and along the 
fission path~\cite{Ryssens23}. The description of deformed nuclear clusters is also of importance for the 
deep layers of the NS inner crust, where a competition between the strong and the electromagnetic interactions 
takes place, leading to exotic nuclear configurations~\cite{Ravenhall83,Hashimoto84}. 
These so-called nuclear pasta phases may look like “gnocchi" (droplets), “spaghetti” (rods), “lasagna” 
(slabs), “bucatini” (tubes), and “Swiss cheese” (bubbles) and are predicted to exist in cold and hot stellar matter. 
Moreover, BSkG models allow for time-reversal breaking and can hence access spin and current densities; the latter 
enter so-called `time-odd terms' in the EDF, which contribute to the total energy of odd and odd-odd nuclei
and are crucial for describing spin- and spin–isospin polarized INM.
Aside from their application to 
$r$-process nucleosynthesis~\cite{Goriely2020}, the BSkG models have shown how relevant their symmetry-breaking approach 
is to interpret new experimental data in different regions of the nuclear chart: 
from mass measurements~\cite{hukkanen2023,hukkanen2023a,Hukkanen22} to charge radii extracted from laser spectroscopy~\cite{maass_inprep,warbinek2024}
and even the interpretation of newly discovered isomers~\cite{stryjczykDiscoveryNewLonglived2024}.


In this work we concentrate on improving the $^1S_0$ pairing channel of our models, which plays an essential role in both 
finite nuclei and dense nuclear matter; pairing contributes to binding energy and stability of the former and leads to the 
emergence of superfluidity and superconductivity in the dense matter of NSs~\cite{Chamel17}. Pairing arises due to an 
attractive interaction between fermions, which allows them to form pairs and condense into a superfluid state below a 
critical temperature~\cite{Bardeen57}. Odd-even effects associated with the pairing interaction directly influence the 
neutron separation energies of finite nuclei, a quantity that defines the nuclear path during the $r$-process neutron 
irradiation, as well as their $Q_\beta$ value that determines the $\beta$-decay half-lives, hence the timescales against 
the production of heavy species during the $r$-process \cite{Arnould07}. Similarly, pairing effects are known to influence the prediction 
of fission probabilities that can affect the recycling and final composition resulting from the $r$-process nucleosynthesis \cite{Goriely15}. 
For these reasons, a careful description of the pairing interaction is needed for a reliable prediction of absolute masses 
as well as mass differences and fission properties influencing reaction and decay rates, hence nucleosynthesis simulations.

Another interesting physical system to the study of the pairing phenomena is the inner crust of a NS, where nuclear clusters 
coexist with superfluid neutrons and possibly superconducting protons in the deepest layers. Because the coherence lengths are typically large~\cite{Okihashi2021}, a consistent and reliable treatment 
of the pairing phenomenon in both clusters and INM is required for an accurate modeling of superfluidity in the inner crust of 
NS~\cite{Chamel2010super}. 
Superfluidity plays a major role in various astrophysical phenomena, such as pulsar frequency glitches~\cite{antonopoulou2022}, NS cooling~\cite{potekhin2020} and oscillations~\cite{Andersson21}. 
A new toolkit has been recently released~\cite{Pecak24} to simulate the local superfluid dynamics in NS, allowing for instance to analyze the motion of an impurity  through the neutron superfluid in the NS inner crust, and extract the effective  mass, determine the critical velocities, study the formation of topological defects and 
dissipation processes. 
The tool is based on time-dependent HFB calculations with extended Skyrme functionals such as the BSkG series and will benefit from a more refined treatment of INM pairing gaps at arbitrary asymmetries in the EDF.

The last model of our series, BSkG3~\cite{Grams23}, describes nucleon pairing at low densities based on a specific prescription built to match the $^1S_0$ pairing gaps in INM as determined by realistic Extended Brueckner-Hartree-Fock (EBHF) calculations~\cite{Cao06b}, rather than relying on the purely phenomenological ansatz used in BSkG1-2~\cite{Scamps21,Ryssens22}. However, this match to EBHF calculations is limited to pure neutron matter (NM) and symmetric nuclear matter (SM), requiring an interpolation scheme for arbitrary asymmetries.
All Brussels models since BSk17~\cite{Goriely09} (with the exception of BSk27~\cite{goriely2013}) adopted a simple prescription that later was found to lead to negative proton pairing gaps at high asymmetries. We modified the interpolation prescription for BSkG3~\cite{Grams23} to avoid this issue. However, we found that the improvement on BSkG3 was not enough to reproduce 
the behaviour of pairing gaps at arbitrary asymmetries as given by Bardeen-Cooper-Schrieffer (BCS) approximation \cite{Zhang_Cao_etal_10}. To address this problem we propose a new entry for the BSkG series, which improves the description of pairing gaps in highly asymmetric environments. This is achieved with BSkG4 by modifying the pairing term of the EDF with an improved interpolation ansatz for the pairing gaps, while keeping the other model ingredients similar to those used for BSkG3, as detailed in Sec.~\ref{sec:massmodel}. BSkG4  description of finite nuclei is discussed in Sec.~\ref{sec:BSkG4} and its predictions for INM properties and astrophysics applications in Sec.~\ref{sec:INM}. Finally, we reserve the last section to our conclusions.


\section{Model construction and $^1S_0$ pairing gaps}
\label{sec:massmodel}

We describe an atomic nucleus using a Bogoliubov type mean-field state, which we
   obtain as the solution to the self-consistent HFB equations based on a Skyrme-type
   EDF. BSkG4 was constructed along the same lines as BSkG3: it is based on an extended form of the Skyrme functional 
   whose parameters resulted from a large-scale fitting protocol featuring thousands of known nuclear masses, hundreds of charge
   radii, reference values of the fission barriers of actinide nuclei and INM properties~\cite{Grams23}. We relied on the efficient and robust
   MOCCa code; its three-dimensional coordinate-space representation imposes little symmetry restrictions on
   the nuclear configurations while providing excellent numerical precision, even for the elongated
   shapes encountered along fission barriers~\cite{Ryssens15,RyssensThesis}.
   We refer the interested reader for more information to Refs.~\cite{Scamps21,Ryssens22,Grams23}.

The novelty of BSkG4 concerns our treatment of the pairing channel away from the limits of NM and SM.
    The pairing terms in our functional are inspired by those of a zero-range pairing interaction acting between like particles
    but with an effective density-dependent strength $g_q(\rho_n, \rho_p)$ ($q = p,n$):
\begin{align}
g_{q}(\rho_n, \rho_p) &=
V_{q}(\rho_n, \rho_p)  \left[
1 + \kappa_q (\nabla \rho)^2 \right]
\, ,
\label{eq:micro_strength}
\end{align}
where the $\rho_n, \rho_p$ are the nucleonic densities and $\rho = \rho_n + \rho_p$ is the total density.
Using the procedure of Ref.~\cite{goriely+2016}, we determine $V_{q}(\rho_n, \rho_p)$ by requiring that the pairing
gaps corresponding to Eq.~\eqref{eq:micro_strength} locally match those of advanced many-body calculations in
homogeneous INM.
The ansatz for the function $g_{q}$ and the regularization of high-momentum divergences 
are the same as for BSkG3~\cite{Grams23}. In particular, this prescription allows for both the bulk and gradient pairing terms to be properly renormalized with the energy cutoff $E_{\rm cut}$. This is crucial for time-dependent HFB calculations~\cite{pecak_2021} for which very high cutoff
values $E_{\rm cut}\sim 100$~MeV have to be considered to ensure the conservation of energy during the evolution (see, e.g., Ref.~\cite{magierski2019nuclear}). 


We rely on the predictions for the $^1S_0$ nucleon pairing gaps in INM of Ref.~\cite{Cao06} based on EBHF with polarization and
self-energy corrections\footnote{Recently we have implemented a new feature in the MOCCa code - an option for the strictly periodic boundary conditions. This allows us to verify that the code reproduces the reference gaps of Ref.~\cite{Cao06} in NM. The detailed methods and results will be reported in a forthcoming paper.}. In practice, we adopt 
parameterizations of those predictions; these are described in detail in~\ref{app:gapfits}.
  Since the EBHF calculations were restricted to NM and SM, we are forced to seek an interpolation
  formula that can reasonably well describe pairing gaps for arbitrary isospin asymmetries $\delta=(\rho_n-\rho_p)/\rho$.
  A reasonable initial step on this road is assuming isospin symmetry, i.e. that $\Delta_n(\rho_n, \rho_p) = \Delta_p(\rho_p, \rho_n)$.
 The first interpolation
  was proposed in Ref.~\cite{Goriely09}:
\begin{align}
 \Delta_q(\rho_n,\rho_p)=(1-|\delta|)\Delta_\mathrm{SM}(\rho)\pm \delta\frac{\rho_q}{\rho}\Delta_\mathrm{NM}(\rho_q)\, ,
\label{eq:GCP09}
\end{align}
where the upper (lower) sign is to be taken for neutrons (protons).
This interpolation was 
adopted in all of the Bogoliubov-Skyrme (BSk) 
parametrizations since BSk17~\cite{Goriely09}, except for BSk27~\cite{goriely2013}. 
However, Eq.~\eqref{eq:GCP09} has the shortcoming that it produces negative pairing gaps
in specific density ranges when $\delta$ was close (but not equal to) -1.
Although one can manually set the gap to zero in such cases,
we avoided this spurious feature for BSkG3~\cite{Grams23} by using instead:
\begin{align}
\Delta_q(\rho_n,\rho_p)=(1-|\delta|)\Delta_\mathrm{SM}(\rho)+|\delta|\Delta_\mathrm{NM}(\rho_q)\, .
\label{eq:GRSGC23}
\end{align}
However, Eq.\eqref{eq:GRSGC23} remains a simple ad-hoc proposal.

\begin{figure}
  \includegraphics[width=\columnwidth]{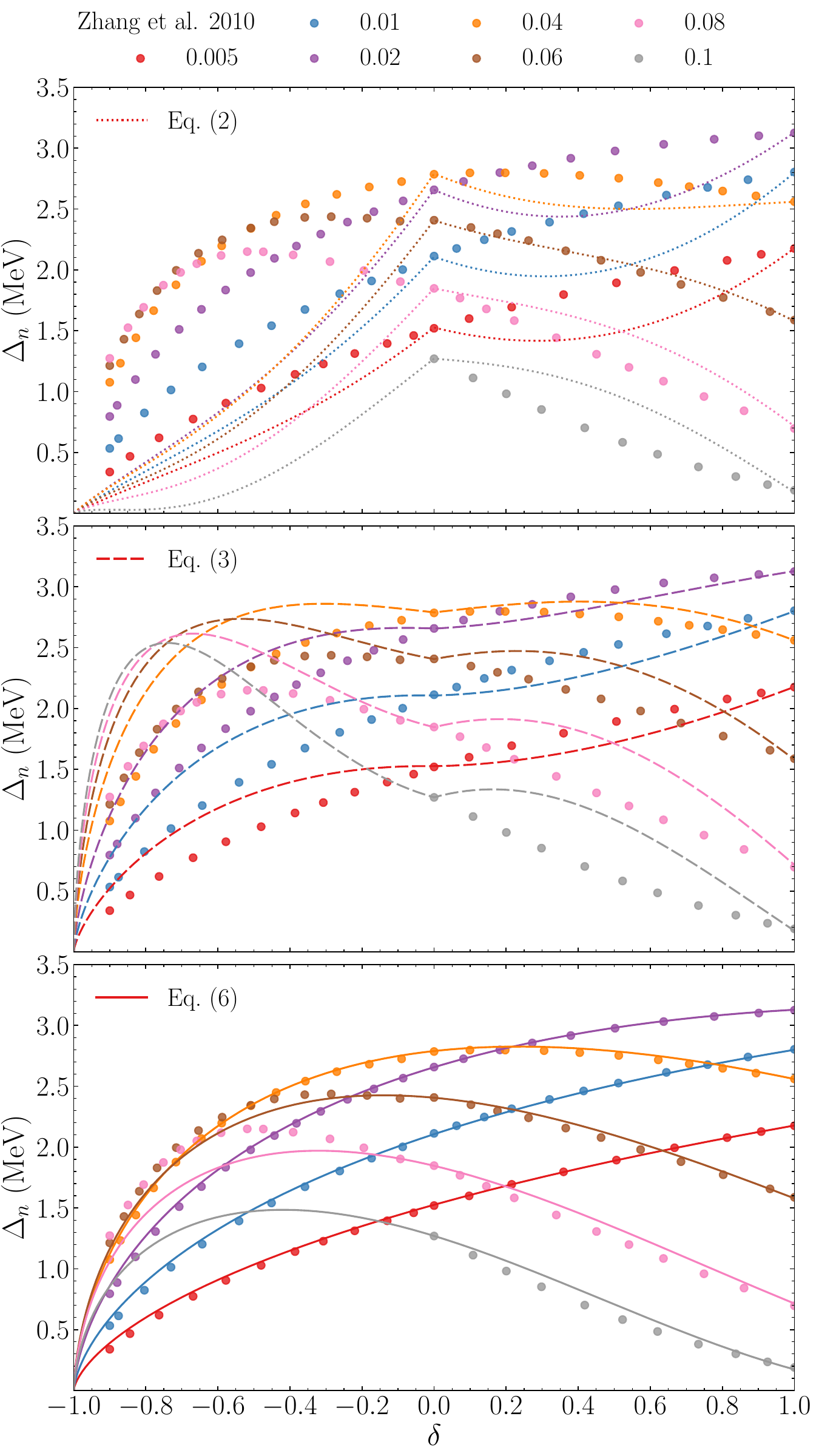}
\caption{$^1S_0$ neutron pairing gaps in asymmetric INM as a function of the isospin asymmetry parameter $\delta$ (the proton pairing gaps are the same if one changes $\delta$ to $-\delta$). 
Points represent the results of Ref.~\cite{Zhang_Cao_etal_10} at different total densities $\rho$ (in fm$^{-3}$).
The dotted lines on the top panel correspond to the interpolation from Eq.~\eqref{eq:GCP09}. The middle panel shows the same quantity in dashed lines for the interpolation scheme of Eq.~\eqref{eq:GRSGC23}. The bottom panel displays in solid lines the new interpolation of Eq.~\eqref{eq:interp_new}.
All three panels interpolate the symmetric and neutron matter gaps from Ref.~\cite{Zhang_Cao_etal_10}.}
\label{fig:gapINM_Zhang}   
\end{figure}

Zhang et al. \cite{Zhang_Cao_etal_10} investigated the isospin dependence of the 
$^1S_0$ pairing gaps in asymmetric INM 
from BCS calculations using the realistic Argonne AV$_{18}$ potential.
Although they did not include polarization and self-energy corrections, their results can still provide 
some guidance. 
As shown in Fig.~\ref{fig:gapINM_Zhang}, neither Eq.~\eqref{eq:GCP09} (dotted lines in the upper panel)
nor Eq.~\eqref{eq:GRSGC23} (dashed lines in the middle panel) can reproduce the results of
Ref.~\cite{Zhang_Cao_etal_10} (this flaw of Eq.~\eqref{eq:GCP09} was already pointed out in Ref.~\cite{Zhang_Cao_etal_10}). 
For this comparison, 
we use analytic representations for $\Delta_\mathrm{NM}$ and $\Delta_\mathrm{SM}$ fitted to 
the results from Ref.~\cite{Zhang_Cao_etal_10} in NM and SM respectively. 

This disagreement motivated us to search for a better interpolation. 
As was argued in Ref.~\cite{Zhang_Cao_etal_10}, the isospin dependence of the pairing gaps mainly comes from the isospin splitting of the effective mass $m^*_q$. Therefore we adopt the 
ansatz 
\begin{align}
\Delta_q(\rho_n,\rho_p)=\Delta_\mathrm{NM}(\rho_q)\cdot \exp\left(\frac{C_1(\rho)}{m^*_q(\rho,\delta)}+C_2(\rho)\right).
\end{align}
Then we can determine the functions $C_1 (\rho)$ and $C_2(\rho)$ by
requiring $\Delta_n(\rho,0)=\Delta_p(0,\rho)=\Delta_\mathrm{NM}(\rho)$ and $\Delta_q(\rho/2,\rho/2)=\Delta_\mathrm{SM}(\rho)$.
Furthermore, the effective mass for the extended Skyrme form we use can be expressed as
\begin{align}
\dfrac{m^*_q(\rho,\delta)}{m_q}= \left[ 1+a(\rho)\pm b(\rho)\delta \right]^{-1}\, ,
\end{align}
with upper (lower) sign for $q=n(p)$~\cite{Chamel09}; $a(\rho),\, b(\rho)$ are functions of the total density only.
After some algebra, we find
\begin{align}\label{eq:interp_new}
\Delta_q(\rho_n,\rho_p)&=\Delta_{\rm NM}(\rho_q) \left[\frac{\Delta_{\rm SM}(\rho)}{\Delta_{\rm NM}(\rho/2)}\right]^{\left(1\pm\delta\right)}\, ,
\end{align}
with the lower (upper) sign for neutrons (protons). We show in the bottom panel of Fig.~\ref{fig:gapINM_Zhang} that Eq.~\eqref{eq:interp_new} is much better at capturing the
isospin dependence of the predictions of Ref.~\cite{Zhang_Cao_etal_10}.

\begin{figure}
  \includegraphics[width=\columnwidth]{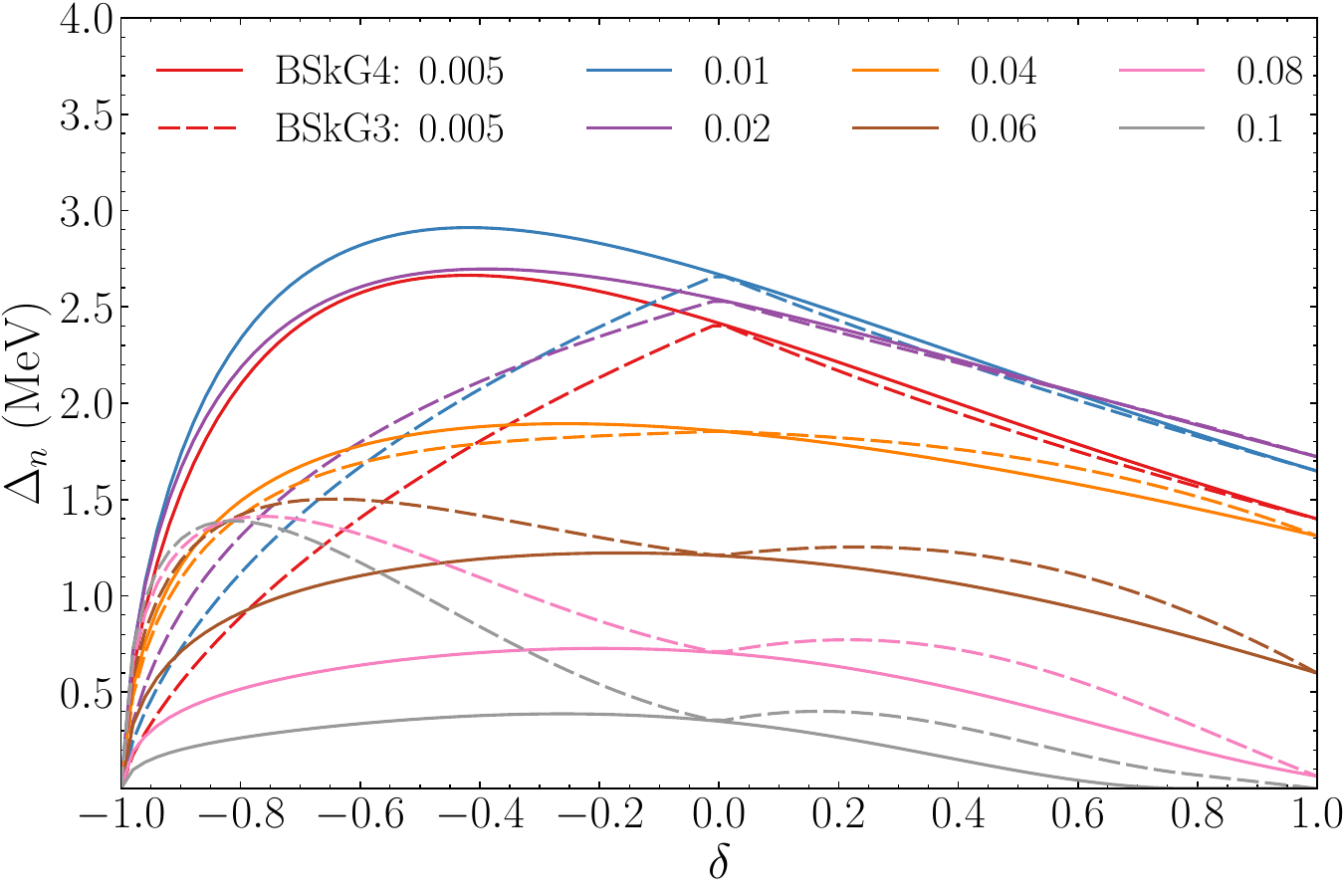}
\caption{$^1S_0$ neutron pairing gaps in asymmetric INM for the BSkG3 (dashed lines) and BSkG4 (solid lines)
        parametrizations. Colors correspond to different total densities $\rho$ (in fm$^{-3}$).
}
\label{fig:gapINM_Cao}   
\end{figure}

We now apply the new  formula to the EBHF predictions of Cao et al.~\cite{Cao06}; Fig.~\ref{fig:gapINM_Cao}
compares the results with those produced by the interpolation used for BSkG3 (Eq.~\ref{eq:GRSGC23}). Our new approach generally predicts
smaller values of $\Delta_n$ in neutron-rich matter. We will see in Sec.~\ref{sec:BSkG4} that this results in somewhat smaller pairing gaps 
in heavy nuclei compared to our previous interpolation. The largest differences, however, arise in proton-rich
matter: $\Delta_n$ is enhanced at densities $\rho\lesssim 0.04$~fm$^{-3}$ and suppressed at higher densities.
As will be discussed in Sec.~\ref{sec:INM}, the latter property has important consequences for proton superconductivity in
the NS interior. Finally, the new interpolation removes the kink at $\delta=0$ and renders the derivation of thermodynamic quantities easier.

%

\section{Properties of atomic nuclei}
\label{sec:BSkG4}


\begin{table}[t]
\caption{
The BSkG4 parameter set: seventeen parameters determining the self-consistent
mean-field energy $E_{\rm HFB}$, three to the pairing 
functional, and nine determining the correction energy 
$E_{\rm corr}$ (see Ref.~\cite{Grams23} for details on $E_{\rm HFB}$ and $E_{\rm corr}$). For comparison, we include the values of the BSkG3 parameter
set~\cite{Grams23}. Note that instead of parameter $x_2$ we list the 
values of the product $x_2 \, t_2$.
}
\begin{tabular}{l|d{6.8}|d{6.8}|d{6.8}}
\hline
    Parameters                            & {\rm BSkG3}  &  {\rm BSkG4}  \\
\hline
$t_0$         [MeV fm$^3$]               & -2325.35     &  -2325.45   \\
$t_1$         [MeV fm$^5$]               &  749.82      & 731.84   \\
$t_2$         [MeV fm$^5$]               &  0.01        &   0.01 \\
$t_3$         [MeV fm$^{3 + 3\alpha}$]   & 14083.45     & 14092.79   \\
$t_4$         [MeV fm$^{5 + 3\beta}$]    & -498.01      & -476.32   \\
$t_5$         [MeV fm$^{5 + 3\gamma}$]   & 266.52       & 271.19   \\
$x_0$                                    &  0.558834    & 0.549106   \\
$x_1$                                    &  2.940880    & 2.97317   \\
$x_2 t_2$     [MeV fm$^5$]               & -432.256954  & -431.435904   \\
$x_3$                                    & 0.628699     & 0.618431   \\
$x_4$                                    &  5.657990    & 5.87636    \\
$x_5$                                    &  0.396933    & 0.353345    \\
$W_0$         [MeV fm$^5$]               & 119.735      & 122.206   \\
$W_0'$        [MeV fm$^5$]               & 78.988       & 79.840   \\
$\alpha$                                 & 1/5          &  1/5     \\
$\beta$                                  & 1/12         &  1/12  \\
$\gamma$                                 & 1/4          &  1/4  \\
\hline
$\kappa_{n}$ [fm$^8$]                    & 123.20       &   123.20  \\
$\kappa_{p}$ [fm$^8$]                    & 129.07       &   129.07 \\
$E_{\rm cut}$ [MeV]                      & 7.961        &   7.919 \\
\hline
$b$                                      &  0.810       &  0.905  \\
$c$                                      &  7.756       &  6.764  \\
$d$                                      &  0.289       &  0.234  \\
$l$                                      &  5.499       &  1.787   \\
$\beta_{\rm vib}$                        &  0.827       &  0.866  \\
\hline
$V_W$         [MeV]                      & -1.716       & -1.411   \\
$\lambda$                                & 437.20       & 560.00   \\
$V_W'$        [MeV]                      & 0.502        & 0.531   \\
$A_0$                                    & 37.801       &  38.174  \\
\hline
\end{tabular}
\label{tab:param_skyrme}
\end{table}

\begin{table}[h]
\centering
\caption{Root-mean-square 
         ($\sigma$) and 
         mean  ($\bar \epsilon$) deviations between experiment and predictions for the BSkG3 and BSkG4 models. 
         The first block refers to the nuclear ground-state properties and the second one to fission 
         properties. More specifically, these 
         values were calculated with respect to 2457 known masses ($M$)~\cite{AME2020} 
         of nuclei with $Z$, $N \geq 8$, the subset of 299 known masses $M_{nr}$ of neutron-rich nuclei with $S_n \le 5$~MeV, 
         the 1784 (2109) five-point neutron (proton) gaps $\Delta^{(5)}_{n}$ $(\Delta^{(5)}_{p})$ ,
         2309 neutron separation energies 
         ($S_n$), 2173 $\beta$-decay energies ($Q_\beta$), 810 measured charge radii 
         ($R_c$) \cite{Angeli13}, 45 empirical values for primary ($E_{\rm I}$)
         and secondary ($E_{\rm II}$) fission barrier heights~\cite{Capote09} 
         and 28 fission isomer excitation energies ($E_{\rm iso}$) of actinide nuclei~\cite{Samyn04}. The first line gives the model error \cite{Moller88} on all the measured masses. } 
\begin{tabular}{l|d{6.8}|d{6.8}|d{6.8}}
\hline
\hline
Results                       &   {\rm  BSkG3}  &  {\rm  BSkG4} \\ 
\hline
$\sigma_{\rm mod}(M)$ [MeV]           &    0.627    &  0.629 \\ 
$\sigma(M)$ [MeV]                     &   0.631     &  0.633 \\ 
$\bar \epsilon (M)$ [MeV]             &  +0.080     &  +0.104  \\ 
$\sigma(M_{\rm nr})$ [MeV]            &    0.660    & 0.672  \\ 
$\bar \epsilon (M_{\rm nr})$ [MeV]    &   -0.011    & +0.037  \\ 
$\sigma(\Delta_n^5)$ [MeV]            &   0.300     &  0.290  \\ 
$\bar \epsilon (\Delta_n^5)$ [MeV]    &   -0.074    & -0.041   \\ 
$\sigma(\Delta_p^5)$ [MeV]            &   0.428     &  0.427  \\ 
$\bar \epsilon (\Delta_p^5)$ [MeV]    &  +0.005     & +0.005   \\ 
$\sigma(S_n)$ [MeV]                   &   0.442     &  0.402  \\ 
$\bar \epsilon (S_n)$ [MeV]           &  +0.009     & +0.010   \\ 
$\sigma(Q_\beta)$ [MeV]               & 0.534       & 0.493  \\ 
 $\bar \epsilon (Q_\beta)$ [MeV]       &  +0.021     & +0.026    \\ 
$\sigma(R_c)$ [fm]                    &  0.0237     &  0.0246  \\ 
$\bar \epsilon (R_c)$ [fm]            &  +0.0006    & +0.0006   \\ 
\hline
$\sigma(E_{\rm I})    $  [MeV]        &    0.33    & 0.36   \\
$\bar{\epsilon}(E_{\rm I})  $  [MeV]  &   +0.06    & -0.02     \\
$\sigma(E_{\rm II})   $  [MeV]        &    0.51    &  0.53  \\
$\bar{\epsilon}(E_{\rm II}) $  [MeV]  &   +0.01    &  -0.02   \\
$\sigma(E_{\rm iso})  $  [MeV]        &   0.36     &  0.33   \\
$\bar{\epsilon}(E_{\rm iso})$  [MeV]  &  -0.05     & +0.05    \\
\hline
\hline
\end{tabular}
\label{tab:rms}
\end{table}

BSkG4 shares all model ingredients with BSkG3, with the exception of the interpolation of pairing gaps in INM. 
Both models have therefore the same number of 29 parameters. The BSkG4 parameters were refitted using the same fitting protocol as for BSkG3 and starting with BSkG3 values, resulting in similar or identical values, as shown in Table 1. 
Hence, differences between both models arise mainly from the novel approach to the pairing channel.
We refer the reader to Refs.~\cite{Scamps21,Ryssens22,Grams23} for more information on the role of all parameters and the fitting protocol.

\subsection{Ground state properties}
\label{sec:groundstate}

Table~\ref{tab:rms} showcases BSkG4 ability to globally describe the properties of atomic nuclei. The tables first block concerns ground state 
properties. We list the mean ($\bar{\epsilon} =$ experiment - model) and rms ($\sigma$) deviations of the model compared to AME20 data~\cite{AME2020} for nuclei with $N,Z \geq 8$:
their nuclear masses ($M$), separation energies ($S_n$), five-point neutron and proton gaps $(\Delta^{(5)}_{n/p})$ and $\beta$-decay energies $(Q_{\beta})$; we also show the mean and rms deviations of the charge radii we obtain w.r.t. 810 measured charge radii of Ref.~\cite{Angeli13}\footnote{We exclude the values of Re, Po, Rn, Fr, Ra, Am and Cm isotopes listed in Ref.~\cite{Angeli13}: these were determined from systematics rather than measurements.}. In the second block of
Table~\ref{tab:rms}, we treat the fission properties of actinide nuclei: the mean and rms deviations with respect to all 45 reference values in the RIPL-3 database~\cite{Capote09} for the primary ($E_{\rm I}$) and secondary $(E_{\rm II})$ barriers of $Z\geq 90$ nuclei are listed, as are the mean and rms deviations 
of the excitation energies of 28 known fission isomers among the same set of nuclei.

\begin{figure}
  \includegraphics[width=0.45\textwidth]{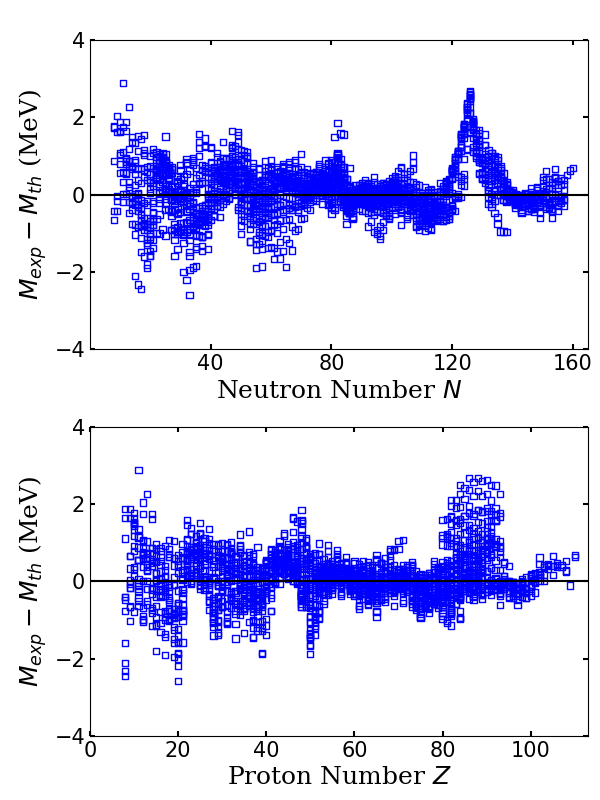}
\caption{Differences between experimental \cite{AME2020} and BSkG4 masses as a function of the neutron number (top) and proton number (bottom).}
\label{fig:massexp}   
\end{figure}

\begin{figure}
  \includegraphics[width=0.45\textwidth]{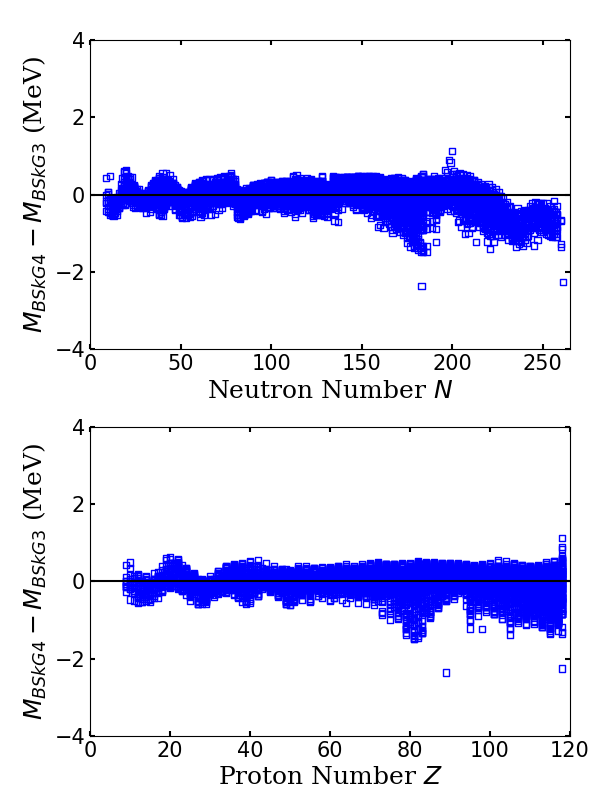}
\caption{Differences between BSkG4 and BSkG3~\cite{Grams23} masses as a function of the neutron number (top) and proton number (bottom) for all the 6946 nuclei with $ 8 \leq Z \leq 118$ lying between the BSkG4 proton and neutron drip lines.}
\label{fig:massdiff}   
\end{figure}

Generally speaking, BSkG4 achieves the same level of global agreement with data as BSkG3: the rms deviations for the masses for the two models are nearly identical, both globally as well as when restricted to the subset of 299 known masses of neutron-rich nuclei with $S_n \leq 5$ MeV. This is reflected too in 
Fig.~\ref{fig:massexp}, which shows the difference between the calculated and experimental masses as a function of neutron number (top panel) and proton number
(bottom panel). 
This figure is comparable to its equivalent for BSkG3 (Fig. 2 in Ref.~\cite{Grams23}) and in particular exhibits the same patterns w.r.t. the location of the largest deviations.
The difference between BSkG4 and BSkG3 atomic masses is shown in Fig.~\ref{fig:massdiff} for all the 6946 nuclei with $8 \le Z \le 118$ lying between the BSkG4 proton and neutron drip lines. Both models produce very similar masses, with a difference typically below 1 MeV. Some differences appear for exotic heavy nuclei, where BSkG4 obtains binding energies predominantly larger than BSkG3.

Nevertheless, BSkG4 performs better for mass differences: 
$\sigma(S_n)$ and $\sigma(Q_{\beta})$ are reduced by 9\% and 8\%, respectively. Although these improvements appear modest, they establish BSkG4 as the most performant of all BSk and BSkG models for these observables; we are not aware of any other set of microscopic calculations that can match this level of accuracy for all $S_n$ and $Q_{\beta}$.
On the other hand, BSkG4 performs slightly worse than BSkG3 for the rms deviations of charge radii and the primary fission barriers of actinide nuclei; even if $\sigma(R_c)$ and $\sigma(E_I)$ increased by 4\% and 9\% respectively.
The mean deviation of the $R_c$ however remains the same level as BSkG3, and $\bar{\epsilon}(E_I)$ is three times lower than for BSkG3.
In summary, BSkG4 retains a global description of both properties that is highly competitive. 

The rms deviations of the five-point pairing gaps in Table~\ref{tab:rms} do not immediately point to an improvement in our description of nuclear pairing properties: $\sigma(\Delta_n^{(5)})$ was improved by only 3\% while $\sigma(\Delta_p^{(5)})$ is essentially unchanged. The mean deviation of the neutron pairing gaps indicates a strong improvement though: $\bar{\epsilon}(\Delta^{(5)}_n)$ was nearly cut in half. As foreshadowed in Sec.~\ref{sec:massmodel}, what produces these changes is the modification to our effective 
neutron pairing strength in heavy nuclei: the new interpolation formula produces smaller neutron pairing gaps for modest positive asymmetry $\delta$
across the range of densities most relevant to pairing in finite nuclei. The improvement becomes more evident when one restricts the comparison to nuclei with $A \geq 150$; for that data set, $\epsilon(\Delta^{(5)}_n) = -0.082$ MeV for BSkG3 and only $-0.029$ MeV for BSkG4. However, there is no global improvement for the proton five-point gaps.

\begin{figure}
  \includegraphics[width=0.45\textwidth]{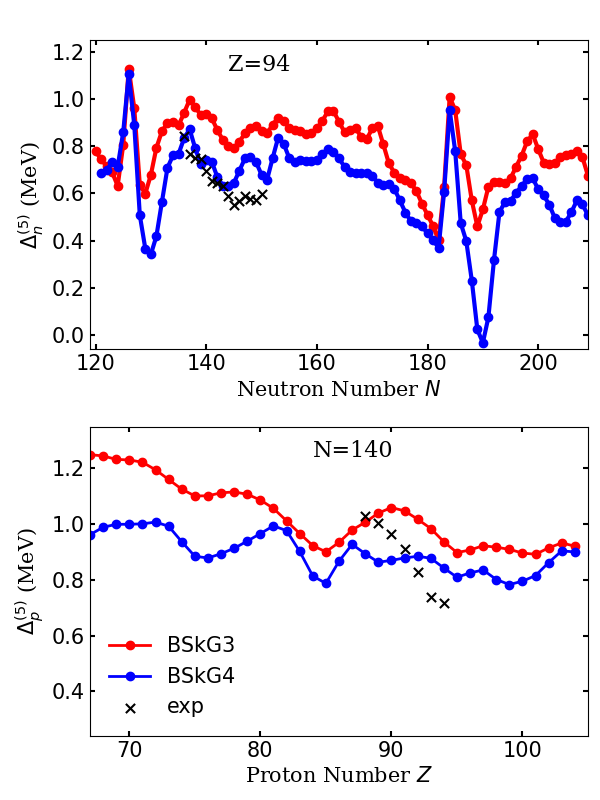}
\caption{A comparison of five-point gaps $\Delta^{(5)}_q$ obtained with BSkG3 and BSkG4 to experimental values. Top panel: neutron values along the Pu isotopic chain. Bottom panel: proton values along the $N=140$ isotonic chain.
}
\label{fig:d5}   
\end{figure}

We illustrate  
these features
with two local examples: the top (bottom) panel of Fig.~\ref{fig:d5} shows the neutron (proton) five-point difference along the 
Pu isotopic ($N=140$ isotonic) chain. The $\Delta^{(5)}_n$ predicted by BSkG4 are significantly lower than those of BSkG3 along the entire $Z=94$ chain; for the small range of neutron numbers where experimental data is available, BSkG4 performs markedly better than its predecessor even if the experimental trend is not perfectly reproduced near $N\sim 150$. On the other hand, the match between the calculated $\Delta^{(5)}_p$ values and experiment along the $N=140$ chain is qualitative at best for both models, even if BSkG4 arguably grasps the average of the experimental values somewhat better. Changing the interpolation formula is clearly not sufficient to cure such gross mismatches between experimental and calculated 
trends; these likely indicate a more fundamental issue with the single-particle structure predicted for these isotopes by both models~\cite{Bender00b}.

\begin{figure}
  \includegraphics[width=0.5\textwidth]{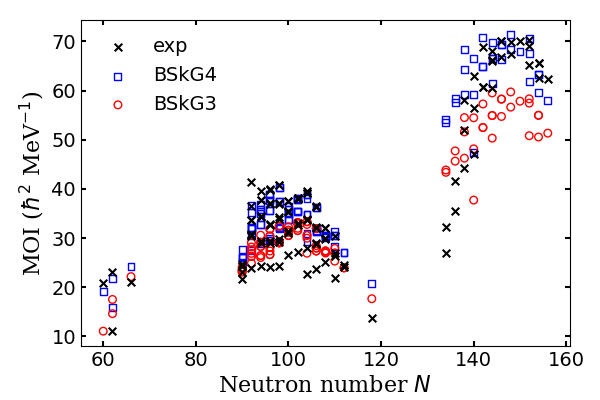}
\caption{Calculated moments of inertia (MOI) as a function of neutron number for BSkG3 (red circles) and BSkG4 (blue squares). Experimental data~\cite{Zeng94,Afanasjev00,Pearson91,Raman01} are shown by black crosses. See text for details.}
\label{fig:MOI}       
\end{figure}

Nuclear pairing impacts other observables besides the binding energy: one example is the rotational moment of inertia (MOI). This observable is sensitive to both pairing and deformation and is crucial to models like ours: it serves as an additional ingredient in the modelling of various collective effects that cannot be captured by simple mean-field calculations, from 
contributions to the binding energy~\cite{Scamps21} to the rotational enhancement of nuclear level densities~\cite{hilaire01}.
Fig.~\ref{fig:MOI} shows the calculated MOIs\footnote{For simplicity, we show the calculated Belyaev MOIs multiplied by 1.32; we have confirmed in Ref.~\cite{Ryssens22} that this old recipe of Ref.~\cite{Libert99} is an acceptable substitute for the determination of Thouless-Valatin MOIs based on full-fledged cranking calculations. } for both models as well as experimental data; the comparison concerns all even-even nuclei with known $2_1^+$ excitation energies below 150 keV. Although it performs reasonably well for medium-mass nuclei, BSkG3 systematically underestimates the MOI for nuclei in the actinide region as already noted in Ref.~\cite{Grams23}, a deficiency shared with both of its predecessors~\cite{Scamps21,Ryssens23}. By reducing the neutron pairing strength, the moments of inertia for heavy nuclei 
trend upwards such that BSkG4 now provides a globally acceptable description of rotational MOIs.  

Finally, we comment on a striking feature of the new model: the prevalence of ground state octupole deformation. When including this degree of freedom, 
   245 nuclei gained more than 500 keV compared to a reflection-symmetric calculation as opposed to just 196 isotopes with BSkG3. The largest energy gain is consistently predicted for $^{286}$Bk, but the BSkG4 value (3.61 MeV) is much larger than the BSkG3 one (2.91 MeV).
   This is a direct result of the new pairing treatment: the smaller pairing strength ensures less competition for (weak) shell effects near the octupole magic numbers to create static ground state reflection asymmetry~\cite{Butler96,Ryssens2019b,Chen21}.

\subsection{Fission properties of BSkG4}
\label{sec:fission}

\begin{figure*}
  \includegraphics[width=\textwidth]{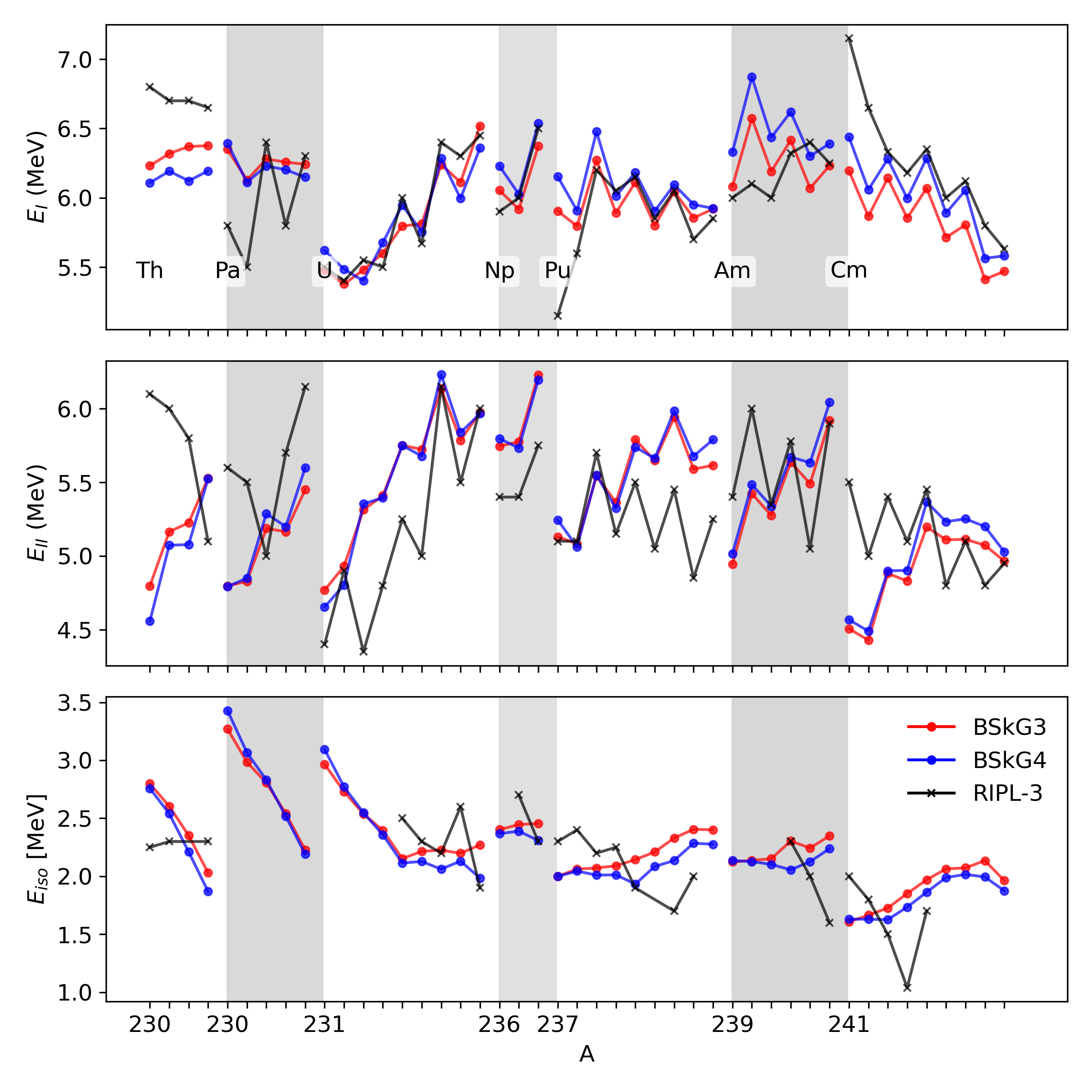}
\caption{Energies of the primary (upper panel) and secondary (middle panel) fission barriers and fission isomer (lower panel) for the 45 actinides in the RIPL-3 database obtained with BSkG3 (red dots) and BSkG4 (blue dots), along with the available reference values (black crosses) \cite{Capote09}. The grey stripes help to differentiate between the isotopic chains shown.}
\label{fig:barriers_BSkG3-4.png}       
\end{figure*}

In addition to the mean and rms deviations shown in Table~\ref{tab:rms}, this section analyzes the differences between BSkG3 and BSkG4 in terms of fission properties. Fig.~\ref{fig:barriers_BSkG3-4.png} displays the primary and secondary fission barriers (upper and middle panels, respectively) as well as the isomer excitation energy (lower panel) for the $Z \ge 90$ nuclei with known empirical barriers included in the RIPL-3 database \cite{Capote09}. 
The numerical details of the calculations performed with the new model are the same as those in the previous work on BSkG2~\cite{Ryssens23}.

In the upper panel of Fig.~\ref{fig:barriers_BSkG3-4.png} we see that BSkG4 predicts slightly higher primary barriers for most nuclei, compared to BSkG3. This is directly attributable to the new pairing treatment: BSkG3 predicted lower primary barriers due to an overestimation of pairing effects. The differences become larger as the mass increases. for instance, Am and Cm chains present more deviations in comparison with Pa or U.

A similar pattern is found for the secondary barriers (middle panel of Fig.\ref{fig:barriers_BSkG3-4.png}).  With the exception of the Th chain, BSkG4 generally produces higher values than BSkG3, with these differences becoming more pronounced as the neutron number increases. Although in some isotopic chains (e.g., Th or Cm) BSkG4 predictions deviate more from the reference values compared to BSkG3, the overall level of agreement remains comparable (see rms deviations in Table~\ref{tab:rms}).

In the case of the isomer excitation energy, BSkG4 systematically predicts lower values. With the reduced pairing content, the shell effects become more pronounced compared to the previous interaction, leading to deeper wells along the fission paths.

Overall, both models predict the same general topography for the actinide barriers. While the deviation between them 
is notable, it remains relatively small compared to the range of calculated values reported in the literature 
~\cite{Ryssens23}. Specifically, the mean and rms deviations between BSkG4 and BSkG3 for the 45 nuclei with $Z \ge 90$ are $\bar{\epsilon}=0.08$~MeV  and $\sigma=0.16$~MeV, respectively, for the primary barrier, 0.03~MeV and 0.09~MeV for the secondary barrier and -0.07~MeV and 0.11~MeV for the isomer excitation energy. Hence, on average, BSkG4 barriers are higher than those of BSkG3 with deviations of the order of 0.1 MeV.

\begin{figure}
  \includegraphics[width=0.48\textwidth]{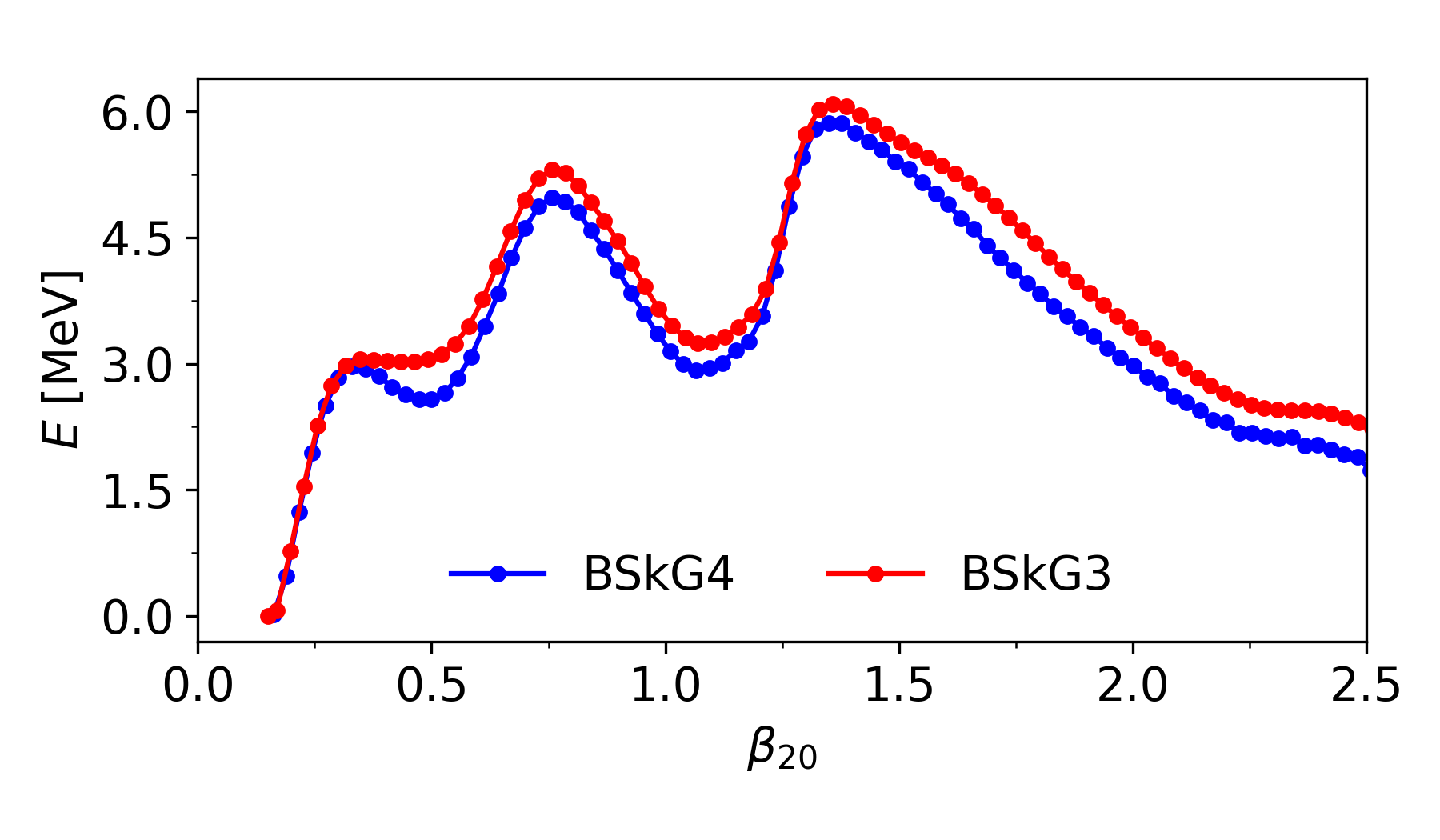}
\caption{Minimum energy fission paths of $^{260}$U as a function of $\beta_{20}$ obtained with BSkG3 and BSkG4.}
\label{fig:U260_fission_paths}       
\end{figure}

To rapidly test if such deviations between BSkG3 and BSkG4 may grow with isospin, 
we also analyzed the neutron-rich isotope $^{260}$U. Fig. \ref{fig:U260_fission_paths} displays the projected 1-dimensional minimum-energy fission path as a function of $\beta_{20}$ for both BSkG3 and BSkG4 parameterizations. To account for triaxial effects, we computed the energy surface in the $(\beta_{20},\beta_{22})$ plane by imposing constraints on both $q_{20}$ and $q_{22}$. Once the surface was defined, we used the PyNEB code~\cite{Flynn2022} to trace the minimum energy (one-dimensional) path from the ground state to the fission configuration.

In contrast to the trend observed in the U chain in Fig.~\ref{fig:barriers_BSkG3-4.png}, BSkG4 predicts slightly smaller primary and secondary barriers. In addition, BSkG4 shows a small initial hump around $\beta_{20} \approx 0.3$, whereas BSkG3 exhibits a flat region at this deformation. Despite these differences, both paths exhibit a similar overall topology with only minor deviations. The refit of the rotational correction parameters plays a significant role: increasing its value (see Table~\ref{tab:param_skyrme}) mitigated the effect of higher rotational MOI induced by BSkG4 lower pairing. 

Although we focused on a static description of fission, the predictions of BSkG4 do not differ significantly from those of BSkG3. The level of agreement with empirical values for actinides is similar, and for the specific case of the exotic neutron-rich $^{260}$U, the static fission paths show only small differences. The generalisation of this statement to all exotic neutron-rich nuclei and the impact of the larger MOI predicted by BSkG4
on the dynamical description of the fission path (where the action is minimized rather than the energy) will be explored in a future work.

\section{Nuclear matter properties and astrophysics applications}
\label{sec:INM}

\subsection{Infinite nuclear matter properties}

We present in Table~\ref{tab:nucmatter} the properties of INM at saturation 
for BSkG3\cite{Grams23} and BSkG4. We show, from top to
bottom, the Fermi momentum $k_{F}$, the saturation density $\rho_{\sat}$, the energy per particle of symmetric matter at saturation
$a_v$, the symmetry energy coefficient $J$, the slope of the symmetry energy 
$L$, the isoscalar (isovector) effective mass $M_s^*/M$ ($M_v^*/M$), the 
compressibility modulus $K_v$, the isovector component of the compressibility 
coefficient $K_{\rm sym}$, the skewness parameter $K^{\prime}$ and the 
Landau parameters $G_{0}$ and $G_0^{\prime}$ of symmetric INM.
For BSkG4, we fix $J = 31$ MeV, as previously done for BSkG3, since we can achieve a stiff symmetry energy thanks to the extended Skyrme form~\cite{Grams23}.
Note that both models present similar INM properties by construction, since we fitted BSkG4 starting from BSkG3. Both models present INM properties at saturation within the current uncertainties~\cite{Margueron2018a}.

\begin{table}[t!]
\centering
\caption{INM properties for the BSkG3~\cite{Grams23} and BSkG4 parameterisations. See Refs.~\cite{Goriely16,Margueron02,Chamel10} 
         for the various definitions. }
\tabcolsep=0.01cm
\begin{tabular}{l *{3}{d{6.5}} }
\hline\noalign{\smallskip}
Properties             & {\rm BSkG3 } &  {\rm BSkG4 }    \\
\hline
$k_F$~[fm$^{-1}$]          & 1.3270   & 1.3270  \\
$\rho_{\sat}$~[fm$^{-3}$]  & 0.15784  & 0.15784 \\
$a_v$~[MeV]                & -16.082  & -16.079 \\
$J$~[MeV]                  & 31.000   & 31.000 \\
$L$~[MeV]                  & 55.061   & 55.674 \\
$M_s^*/M$                  & 0.85860  & 0.86046 \\
$M_v^*/M$                  & 0.72171  & 0.72748 \\
$K_v$~[MeV]                & 242.56   & 242.57 \\
$K_{\text{sym}}$~[MeV]     & -21.248  & -19.479 \\
$K^\prime$~[MeV]           &  304.17  & 303.94 \\
$G_0$                      & 0.20046  & 0.21786 \\
$G_0^\prime$               & 0.97812  & 0.97888 \\
\hline
\end{tabular}
\label{tab:nucmatter} 
\end{table}

\subsection{Neutron star properties}
\label{sec:NS}

\begin{table}[t!]
\centering
\caption{NS 
properties
obtained with BSkG3 and BSkG4 parameterizations. We show in the first row the radius of the canonical NS of $1.4 M_\odot$; the following three lines give the mass, radius and energy density for the most compact NS. $\mathcal{E}_{\rm causal}$ denotes the energy density for which causality is violated. Next we show the threshold energy density and NS mass for the onset of the dUrca process and the energy density $\mathcal{E}_{\rm pQCD}$ up to which the model is consistent with constraints derived from pQCD (see text for details). }
\tabcolsep=0.01cm
\begin{tabular}{l *{3}{d{6.5}} }
\hline
Properties                                         &   {\rm BSkG3}  & {\rm BSkG4}   \\
\hline
$R_{\rm 1.4}$~[km]                                  & 12.73     &  12.74 \\
$M_{\rm max}$~[$M_{\odot}$]                         & 2.26     &   2.26 \\
$R_{\rm M_{\rm max}}$~[km]                           & 11.10    &  11.09 \\
$\mathcal{E}_{\rm M_{\rm max}}$~[$10^{15}$ g cm$^{-3}$]   &  2.273   & 2.282  \\
$\mathcal{E}_{\rm causal}$~[$10^{15}$ g cm$^{-3}$]       & 2.381     &  2.384 \\
$\mathcal{E}_{\rm dUrca}$~[$10^{14}$ g cm$^{-3}$]     & 7.136     &  7.083 \\
$M_{\rm dUrca}$~[$M_{\odot}$]                     & 1.38     &  1.37 \\
$\mathcal{E}_{\rm pQCD}$~[$10^{15}$ g cm$^{-3}$]    & 2.508     &  2.510 \\
\hline
\end{tabular}
\label{tab:NS}
\end{table}

The NS properties for BSkG3 and BSkG4 are summarized in Table~\ref{tab:NS}.
Following the procedure explained in Ref.~\cite{Grams23}, we compute the EoS of npe$\mu$ matter in $\beta$-equilibrium for the NS core and we account for the NS crust by using the approximate formulas given in Ref.~\cite{Zdunik17} for the NS mass and radius.

The first three rows of Table~\ref{tab:NS} correspond to the radius of the canonical 1.4 $M_\odot$ NS, the maximum mass of NS and its respective radius, respectively. 
Next we show the energy density $\mathcal{E}_{\rm causal}$ of $\beta$-equilibrated matter for which the speed of sound exceeds the speed of light. We remark that $\mathcal{E}_{\rm causal}$ for BSkG4 is higher than the central energy density $\mathcal{E}_{\rm M_{\rm max}}$ of the most massive NS. Therefore the newest model (as BSkG3) does not violate causality.

To check if BSkG4 allows for the onset of the direct Urca (dUrca) process~\cite{Lattimer91} we also show in
Table~\ref{tab:NS} the density and the corresponding NS mass for which the proton fraction inside the star reaches the dUrca threshold. This process is required to interpret the observed thermal luminosity of some NS~\cite{Burgio21,Marino2024,Ho_Pol_Deller_Becker_Burke-Spolaor_2024}.
The BSkG4 and BSkG3 models allow for the dUrca process for NS with a mass above $\sim 1.4$ M$_{\odot}$.

To further test our models, we use results from perturbative quantum chromodynamics (pQCD) calculations: 
assuming only causality and thermodynamic consistency, pQCD predictions at extreme 
densities ( $\gtrsim$ 40 $\rho_{\rm sat}$) impose constraints on the EoS
at lower densities~\cite{Komoltsev22}. We have verified that the 
BSkG4 EoS is consistent with this constraint up to an energy density 
$\mathcal{E}_{\rm pQCD} = 2.510 \times 10^{15}$~g~cm$^3$, i.e. above 
the density of the most massive NS predicted by this model.
BSkG3 also satisfies this constraint. 

\begin{figure}
\includegraphics[width=1.\columnwidth]{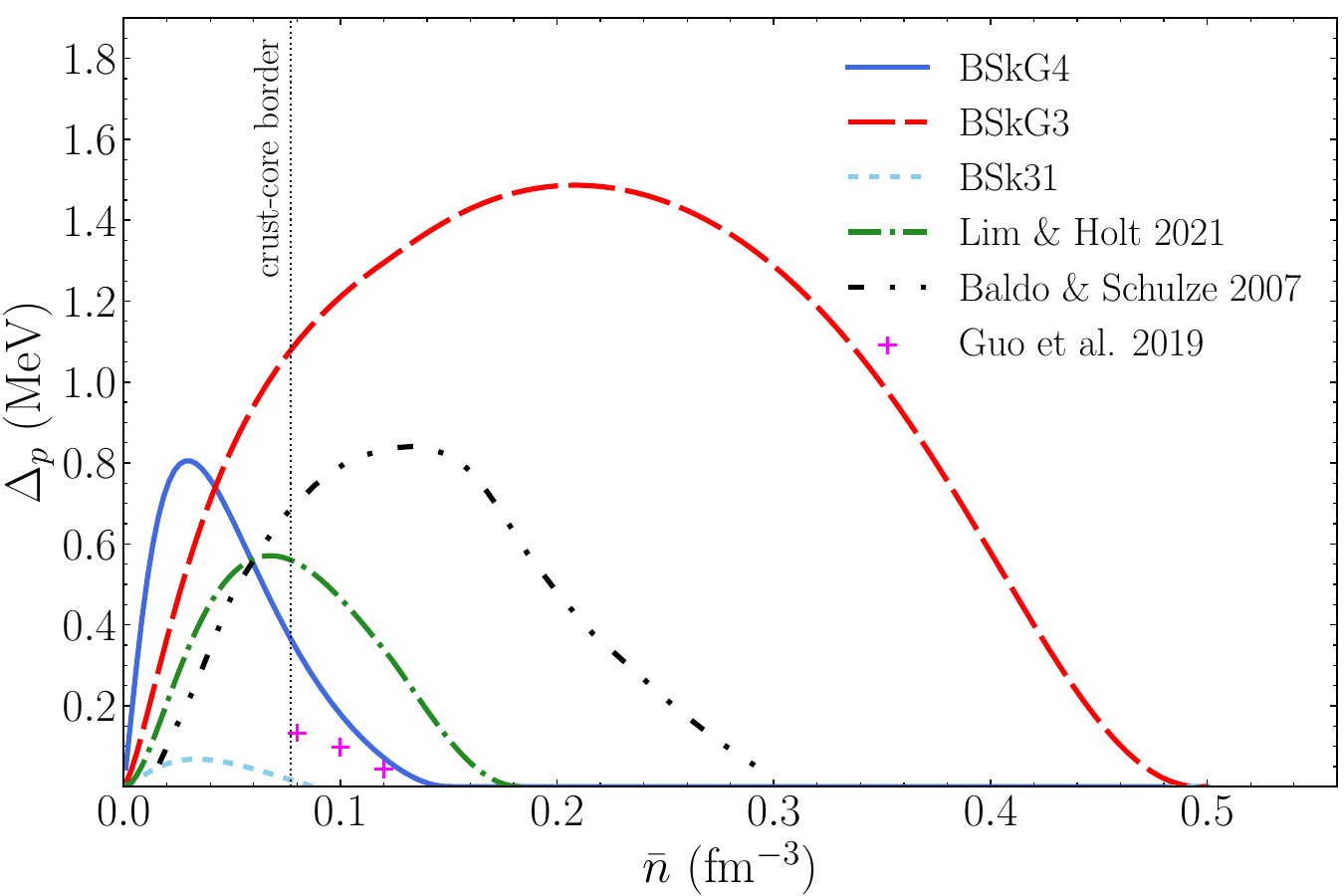}
	\caption{Proton pairing gap in $\beta$-equilibrated infinite npe$\mu$ matter. The blue solid line corresponds to BSkG4, the red dashed line to BSkG3 and the cyan short-dashed line to BSk31. 
 We compare our results with more advanced many-body calculations from Lim \& Holt 2021 (green dash-dotted line)~\cite{Lim21}, Baldo \& Schulze 2007 (black dash-dash-dotted line)~\cite{Baldo07}, and Guo et al. 2019 (magenta crosses)~\cite{Guo19}. 
 The vertical black dotted line marks the approximate density of the crust-core transition in NS. More detail in the text.
 }
 \label{fig:pgap_NS}
\end{figure}

To illustrate the impact of BSkG4 model improvement on NS matter, we plot in Fig.~\ref{fig:pgap_NS} the 
proton pairing gap in $\beta$-equilibrated infinite $npe\mu$ matter. We compare BSkG4 (blue solid line), BSkG3~\cite{Grams23} (red dashed line) and BSk31~\cite{goriely+2016} (cyan short-dashed) with more advanced many-body calculations from: Baldo \& Schulze 2007 (black dash-dash-dotted line)
\cite{Baldo07}, Guo et al. 2019 (magenta crosses) 
\cite{Guo19}, and Lim \& Holt 2021 (green dash-dotted line) obtained using the fitting formula for the n3lo450 plus second-order single-particle energy model from \cite{Lim21} with the proton fraction equal to the one of BSkG4. 
We show the density of crust-core transition with a vertical black line.
In the core of the star, where protons are unbound, their pairing gap within BSkG4 tends closer to the realistic calculations given by \cite{Lim21,Baldo07,Guo19}. 
Even if the spread in the predictions of advanced many-body calculations indicates the uncertainty of the modelling of pairing in NM, the curves of 
BSkG3~\cite{Grams23} and BSk31~\cite{goriely+2016} are clearly too extreme.


In summary, the NS properties 
predicted by BSkG4 are almost identical to the ones obtained with BSkG3. The improvement of BSkG4 appears for properties sensitive to the pairing gaps, as shown in Fig.~\ref{fig:pgap_NS}. 

\subsection{Simulation of $r$-process nucleosynthesis}

\begin{figure}
\includegraphics[width=1.\columnwidth]{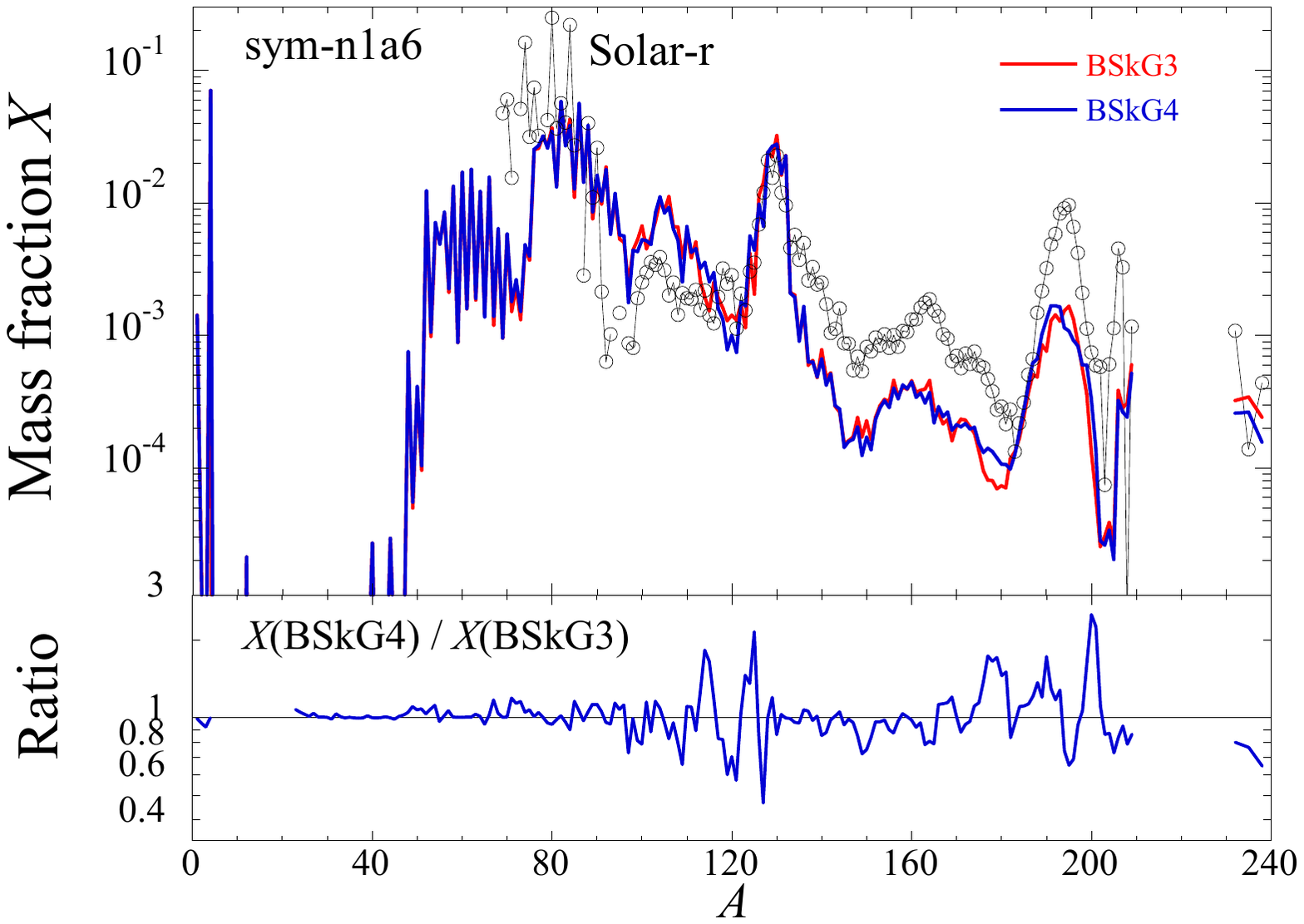}
\caption{Final mass fractions $X$ of the material ejected from a $1.375-1.375~M_{\odot}$ NS-NS binary system of Ref.~\cite{Just23}, obtained with the new BSkG4 and former BSkG3~\cite{Grams23} masses.
The solar system r-abundance distribution  (open circles) from \cite{Goriely99} is shown for comparison and arbitrarily normalised. The lower panel shows the ratio between the predicted BSkG4 and BSkG3 mass fractions for mass fractions $X > 10^{-8}$.
}
\label{fig:rprocess}
\end{figure}

The impact of our new mass model on $r$-process nucleosynthesis is illustrated in Fig.~\ref{fig:rprocess}: it shows the composition of the matter ejected from the specific end-to-end simulation of a $1.375-1.375~M_{\odot}$ NS-NS binary system \cite{Just23}  (the so-called sym-n1a6 model). The final abundance distributions are calculated using the radiative neutron capture and photoneutron emission rates obtained consistently from the BSkG3 and BSkG4 models, but the same $\beta$-decay rates from the calculation of Ref.~\cite{Marketin16}. More details on the $r$-process calculations can be found in Ref.~\cite{Just23}.
Globally, BSkG3 and BSkG4 mass models give rise to abundance distributions that agree with each other and match relatively well the solar system r-distribution for nuclei with $80 \la A \la 130$. Due to the dominant contribution of the comparatively high electron fraction in such a NS merger, nuclei above $A \ga 130$ are underproduced with respect to the solar r-distribution~\cite{Just23}. Although the predicted masses of both models typically differ less than $\sim 1$~MeV (see Fig.~\ref{fig:massdiff}), the local mass differences and odd-even effects triggered by the new pairing description induce variations in the composition of up to a factor of 2; this concerns mainly he vicinity of $A\simeq 120$,  $A\simeq 200$ as well as for production of Th and U. This shows that small modifications of nuclear properties affecting in particular neutron separation energies can still affect local aspects of abundance predictions in a non-negligible way. 

\section{Conclusions}
\label{sec:conclusions}

We presented a new entry in the BSkG series of large-scale nuclear structure models: instead of relying on an ad-hoc recipe, we developed a physically motivated approach to extend the EBHF results of Ref.~\cite{Cao06} that guide our treatment of the EDFs pairing terms to arbitrary asymmetries. Compared to BSkG3, this modification leads to a modestly but systematically improved description of pairing in heavy nuclei and significantly different predictions for the pairing properties of neutron-rich matter in NS that qualitatively match more advanced many-body calculations. 
These improvements do not rely on the introduction of additional parameters nor do they come at the expense of our description of other quantities: the global rms deviations on known masses, charge radii and the fission properties of actinide nuclei are close to those of BSkG3, while the rms deviations of $S_n$ and $Q_{\rm \beta}$ have been lowered beyond those of any other microscopic model known to us.
Furthermore, we performed $r$-process nucleosynthesis simulations with the new mass model. Our results show that the modifications with respect to the previous model have a non-negligible impact on the abundance distribution.


With its more physically motivated interpolation of $^1S_0$ pairing gaps between SM and NM, 
BSkG4 is better suited for modelling the thermal evolution of NS than any of our previous EDFs. 
BSkG4 will also allow for more realistic simulations of the neutron superfluid dynamics in 
NS, which remains poorly understood despite its importance for interpreting various 
astrophysical phenomena related to NS, such as glitches. 

The improvement presented here have led to better description of pairing properties in both finite nuclei and INM. Nevertheless,
the lack of ab initio data across the full range of asymmetries remains a key challenge.
In particular, results that incorporate all relevant many-body
effects, such as medium polarization and self-energy corrections, are sparse. 
We thus encourage further studies
of the isospin dependence of the $^1S_0$ pairing gaps in INM. 
This would significantly advance both the developments of EDFs and their astrophysical applications. 

\begin{acknowledgements}
This work was supported by the Fonds de la Recherche Scientifique (F.R.S.-FNRS) and the Fonds Wetenschappelijk Onderzoek - Vlaanderen (FWO) under the EOS Projects nr O022818F and O000422F. 
This research benefited from computational resources made available on the Tier-1 supercomputer Lucia of the Fédération Wallonie-Bruxelles,
infrastructure funded by the Walloon Region under the grant agreement nr 1117545.
Further computational resources have been provided by the clusters Consortium des Équipements de Calcul Intensif (CÉCI), funded by F.R.S.-FNRS under Grant No. 2.5020.11 and by the Walloon Region. 
 W.R. and S.G. gratefully acknowledge support by the F.R.S.-FNRS. This work also received funding from the Fonds de la Recherche Scientifique (Belgium) under Grant No. PDR T.004320 and IISN 4.4502.19. 
\end{acknowledgements}

\appendix

\section{List of abbreviations}
\label{app:acronyms}

\begin{tabular}{ll}
Abbreviation & Meaning \\
\hline
BSkG  & Brussels-Skyrme-on-a-Grid \\
BCS  & Bardeen-Cooper-Schrieffer \\
dUrca & direct Urca (process) \\
EBHF  & Extended Brueckner-Hartree-Fock \\
EDF   & Energy density functional \\
EoS   & Equation of state \\
GW    & Gravitational waves \\
HFB   &  Hartree-Fock-Bogoliubov \\
INM   & Infinite (homogeneous) nuclear matter        \\
MOI   & Moment of inertia \\
NM    & Pure (infinite) neutron matter \\
NS    & Neutron star  \\
SM  & symmetric (infinite) nuclear matter \\
pQCD  & Perturbative quantum chromodynamics \\
rms   & root-mean-square \\
r-process & rapid neutron-capture process \\
\hline
\end{tabular}

\section{Parameterization of the pairing gaps of Ref.~\cite{Cao06}}
\label{app:gapfits}
For ease of use in MOCCa,
we parametrize the gaps $\Delta_\mathrm{SM}(\rho)$ and $\Delta_\mathrm{NM}(\rho_q)$ as:
\begin{align}\label{eq:fit-neutron-matter-gap}
\Delta_\mathrm{NM}(\rho_q) = \theta(k_m - k_{Fq})\Delta_0\frac{k_{Fq}^2}{k_{Fq}^2 + k_1^2}\,
\frac{(k_{Fq} - k_2)^2}{(k_{Fq} - k_2)^2 + k_3^2}
\end{align}
and
\begin{align}\label{eq:fit-symmetric-matter-gap}
\Delta_\mathrm{SM}(\rho) = \theta(k_m - k_F)\Delta_0
\frac{k_F^2}{k_F^2 + k_1^2}\,\frac{(k_F - k_2)^2}{(k_F - k_2)^2 + k_3^2}\, .
\end{align}
Here $k_{Fq}= (3\pi^2 \rho_q)^{1/3}$, $k_F = (3\pi^2 \rho/2)^{1/3}$ and $\theta$ is the unit-step Heaviside function. 
The parameters $\Delta_0, k_1, k_2, k_3$ and $k_m$ were determined by fitting the results 
of Ref.~\cite{Cao06}; the resulting values are listed in Table~\ref{tab:fits-pairing-gaps}. These fits were performed for the construction of BSk30-31-32; although the parameter values were not provided at the time, the quality of the fit can be judged from Fig.~1 in Ref.~\cite{goriely+2016}.

\begin{table}
\centering
\caption{Parameters for Eqs.~\eqref{eq:fit-neutron-matter-gap}-\eqref{eq:fit-symmetric-matter-gap} of the $^1S_0$ pairing gaps in SM and NM calculated in Ref.~\cite{Cao06}.}
\label{tab:fits-pairing-gaps}
\begin{tabular} {lcc}
\hline
                  &   {\rm SM}      &    {\rm NM} \\ 
\hline
$\Delta_0$ [MeV]  & 11.5586   & 3.37968  \\
$k_1$ [fm$^{-1}$] & 0.489932  & 0.556092 \\
$k_2$ [fm$^{-1}$] & 1.31420   & 1.38236  \\
$k_3$ [fm$^{-1}$] & 0.906146  & 0.327517 \\
$k_m$ [fm$^{-1}$] & 1.31      & 1.38     \\
\hline
\end{tabular}
\end{table}

\section{Explanation of the supplementary material}
\label{app:explanation}

We provide as supplementary material the files \newline 
\textsf{Mass\_Table\_BSkG4.dat} and \textsf{Fission\_Table\_BSkG4.dat}.
The former contains the calculated ground state properties of all nuclei with 
$ 8 \leq Z \leq 118$ lying between the proton and neutron drip lines. The latter
contains the fission barriers and isomer excitation energies as calculated for 
all 45 nuclei with $Z \geq 90$ that figure in the RIPL-3 database. The contents 
of both files follow the conventions of the supplementary files of 
Ref.~\cite{Grams23}. For convenience, we repeat the contents of all columns of both files in 
Tables~\ref{tab:suppl1} and \ref{tab:suppl2}.

\begin{table*}[]
\centering
\begin{tabular}{llll}
\hline
\hline
Column &  Quantity & Units & Explanation \\
\hline
1 & Z & $-$ & Proton number\\
2 & N & $-$ & Neutron number\\
3 & $M_{\rm exp}$ & MeV  & Experimental atomic mass excess \\
4 & $M_{\rm th}$ & MeV   & BSkG4 atomic mass excess \\
5 & $\Delta M$ & MeV      & $M_{\rm exp} - M_{\rm th}$\\
6 & $E_{\rm tot}$ & MeV   & Total energy \\
7 & $\beta_{20}$ & $-$    & \multirow{3}*{Quadrupole deformation} \\
8 & $\beta_{22}$ & $-$    & \\
9 & $\beta_2$  & $-$        & \\
10& $\beta_{30}$ & $-$    & \multirow{2}*{Octupole deformation} \\
11& $\beta_{32}$ & $-$    & \\
12 & $E_{\rm rot}$ & MeV  & Rotational correction \\
13 & $\langle \Delta \rangle_n$ & MeV & Average neutron gap\\
14 & $\langle \Delta \rangle_p$ & MeV & Average proton gap\\
15 & $r_{\rm BSkG4}$ & fm  & Calculated rms charge radius\\
16 & $r_{\rm exp}$  & fm  & Experimental rms charge radius\\
17 &$\Delta r$ & fm   &  $r_{\rm exp}- r_{\rm BSkG4}$  \\
18 & $\mathcal{I}^B$ & $\hbar^2$ MeV$^{-1}$ & Calculated Belyaev MOI.\\
19 & par(p) & $-$ & Parity of proton qp. excitation\\
20 & par(n) & $-$ & Parity of neutron qp. excitation\\
\hline 
\hline
\end{tabular}
\caption{Contents of the \textsf{Mass\_Table\_BSkG4.dat} file.}
\label{tab:suppl1}
\end{table*}

\begin{table*}[]
\centering
\begin{tabular}{lllll}
\hline
\hline
Column &  Quantity & Fission property & Units & Explanation   \\
\hline
1 & Z & & $-$ & Proton number \\
2 & N & & $-$ & Neutron number \\
\hline
3 & $E$          & Inner barrier  & MeV  & Barrier height          \\
4 & $\beta_{20}$ &                & $-$  & Quadrupole deformation  \\
5 & $\beta_{22}$ &                & $-$  &                         \\
\hline
6 & $E$          & Outer barrier  & MeV & Barrier height           \\
7 & $\beta_{20}$ &                & $-$ & Quadrupole deformation   \\
8 & $\beta_{22}$ &                & $-$ & \\
9 & $\beta_{30}$ &                & $-$ & Octupole deformation \\
\hline
10 & $E$         & Isomer         & MeV &  Excitation energy     \\
11 & $\beta_{20}$&                & $-$ &  Quadrupole deformation    \\
\hline 
\hline
\end{tabular}
\caption{Contents of the \textsf{Fission\_Table\_BSkG4.dat} file.}
\label{tab:suppl2}
\end{table*}

\bibliographystyle{spphys}       
\bibliography{ref}   

\end{document}